\documentclass[aip,rsi,reprint,graphicx,groupedaddress]{revtex4-1} 
\usepackage{graphicx,amsmath}

\draft 

\begin{document}


\title{Space- and Time-resolved Resistive Measurement of Liquid Metal 
Wall Thickness} 



\author{S.M.H.~Mirhoseini}
\affiliation{Dept of Applied Physics and Applied Mathematics, 
Columbia University, New York, NY 10027, USA}
\author{F.A.~Volpe}
\email[]{fvolpe@columbia.edu}
\affiliation{Dept of Applied Physics and Applied Mathematics, 
Columbia University, New York, NY 10027, USA}



\date{\today}

\begin{abstract}
In a fusion reactor internally coated with liquid metal, it will be
important to diagnose the thickness of the liquid at various locations in the
vessel, as a function of time, and possibly respond to counteract undesired 
bulging or depletion. The electrical conductance between 
electrodes immersed in the liquid metal can be used as a simple 
proxy for the local thickness. Here a matrix of electrodes is
shown to provide spatially 
and temporally 
resolved measurements of liquid metal 
thickness in the absence of plasma. First a theory is developed for 
$m\times n$ electrodes, and then it is experimentally demonstrated for 
3$\times$1 electrodes, as the liquid stands still or is agitated 
by means of a shaker. 
The experiments were carried out with Galinstan, but are
easily extended to Lithium or other liquid metals.  
\end{abstract}


\maketitle 


\section{Introduction}\label{SecIntro}
Liquid metals (LMs) are attractive low-recycling plasma-facing
materials \cite{MajeskiCDX, MajeskiNF, Kaita, MajeskiLTX} 
that could protect the underlying solid walls of a fusion
reactor from high heat and neutron fluxes \cite{Moir, Abdou, Tabares}. 

Lithium has been frequently used 
\cite{MajeskiCDX, MajeskiNF, Kaita, MajeskiLTX}, 
but Tin, Lithium-Tin alloys, Gallium, the molten salt FliBe and other 
materials are also being considered \cite{Moir, Abdou, Tabares}. 
These materials introduce new diagnostic requirements 
compared with confinement devices featuring solid walls. 
For example, CDX-U and LTX were 
equipped with spectrometers in the visible 
and extreme ultraviolet, with special attention paid to neutral Lithium, Li II 
and Li III lines \cite{Kaita2,MajeskiPoP}. 

An additional diagnostic 
requirement is posed by the very fact that these walls are liquid 
and thus can deform \cite{Narula} under the effect of instabilities, 
turbulence, as a result of non-uniform force fields or currents from the plasma 
\cite{Morley,Jaworski,Ait,TahaPPCF}. 
Deformations are undesired 
for various reasons \cite{TahaPPCF}, therefore they 
need to be monitored as a function of 
space and time, with resolutions of the order of a centimeter (in the poloidal 
and toroidal direction) and 10 ms \cite{TahaPPCF}. 

The sensitivity and precision in the radial direction, 
on the other hand, are dictated by the two 
lengthscales that need to be monitored and preserved. These are the 
distance between the LM  
surface and the last closed flux surface (typically few cm) and the LM 
thickness. The thickness 
can range from sub-millimeter to meters, depending whether the LM 
is only used for its benign plasma-facing properties (low erosion, low 
recycling etc.) or is also meant to attenuate heat and neutrons. 
All things considered, millimeter precision is expected to suffice in most 
cases. 

In a previous work \cite{TahaPPCF} we had shown that, 
quite simply, and on the net of small corrections,  
the electrical conductance between two electrodes immersed in the LM 
scales linearly with the local LM thickness. 
Resistive measurements were used to infer the LM thickness 
in a single location and at a single time \cite{TahaPPCF}.
 
Here, after briefly describing the experimental setup (Sec.\ref{SecSetup}),  
we extend the measurements to 
multiple locations, requiring matricial formalism (Sec.\ref{SecSpace}), 
and to multiple times (Sec.\ref{SecTime}).

\section{Experimental setup}\label{SecSetup}
Here we recapitulate an earlier description of the setup \cite{TahaPPCF} and 
report recent improvements. 

The setup features a container filled for 5-25 mm with a low melting
point  (10$^{\rm o}$C) eutectic alloy of Gallium, Indium and Tin
called Galinstan.  This is about as good an electric conductor as
Lithium  (17\% and 16\% of copper, respectively). 

Embedded in the container are  3$\times$4 copper electrodes of 2 mm diameter 
and various lengths, for comparison (1, 16 and 25 mm, 
each with its own advantages and disadvantages \cite{TahaPPCF}).  
Adjacent electrodes are spaced by 25
mm in one direction and 15 mm in  the other. 

The electrodes are
connected to adjustable current-sources as  well as to voltmeters
referenced to ground. A shunt resistor is connected in series 
with the current-source, to measure the electrode current. As of recently, 
voltage and current signals are digitized at up to $10^5$ KSa/s 
and digitally filtered from high-frequency noise (typically $f<$500 Hz). 
A LabView interface analyzes these data and returns the electrical 
resistance and LM thickness between each pair of electrodes, in real time 
(typically every 10-100 ms).

\section{Space-resolved measurements}\label{SecSpace}
\subsection{Theory}
Consider $m\times n$ electrodes, evenly spaced in the $x$ and $y$ 
direction, at distances $dx$ and $dy$, respectively, between adjacent 
electrodes.  
The electrodes are connected to individual power-supplies. 
These can inject or extract 
current in the LM, in the $z$ direction orthogonal to the $xy$ plane. 
However, no charge is accumulated, and $\nabla \cdot {\bf j}$=0. 
That is, Kirchhoff's law
applies: the sum of all currents emitted or collected by an electrode is zero. 
The convention is adopted here 
that emitted currents are positive; collected currents are negative. 
In general, there are five such currents for each 
electrode: four in the $xy$ plane, pointing at adjacent electrodes,   
and one in the $z$ direction. 
Boundary or corner electrodes, on the other hand, only connect to 
three or two adjacent electrodes and one power supply. 

Currents can obviously flow from or to {\em any} other 
electrode, not necessarily adjacent. Nonetheless, it is not necessary to 
model the system as a complicated network where all electrodes are directly 
connected to each other, forming a total of $mn(mn-1)/2$ connections. 
Instead, a Cartesian grid of $2mn-m-n$ resistors suffices.  
In this representation, each electrode is directly connected to only four 
adjacent electrodes via resistors, indicative of the LM thickness in 
between. Currents can flow from one electrode to another along several 
different routes on this Cartesian grid; the total resistance between 
two remote electrodes can be calculated by repetead application of simple 
sum rules for resistors in series or parallel. The total resistance 
between adjacent electrodes, on the other hand is, with good approximation, 
that of the very resistor that directly connects them. 

Let us call $I_{i,j}$ the current emitted by electrode $i,j$ and directed 
at the power supply. Let $V_{i,j}$ denote the potential of electrode $i,j$ 
relative to some ground reference. 
The current emitted from electrode $i,j$ to electrode $i,j+1$ will be 
proportional to the electric field 
$E_y=(V_{i,j+1}-V_{i,j})/dy$, to conductivity $\sigma$, and to the 
cross-sectional area $h_{i,j+1/2} dx$, where $h_{i,j+1/2}$ is the LM height 
in the midpoint between the two electrodes. After similar 
considerations for the other electrodes in the stencil, Kirchhoff's law writes:
\begin{equation}
\begin{split}
  \frac{I_{ij}}{\sigma} 
              =& \frac{V_{i,j}-V_{i,j+1}}{dy} h_{i,j+\frac{1}{2}} dx +
                 \frac{V_{i,j}-V_{i,j-1}}{dy} h_{i,j-\frac{1}{2}} dx \\
              +& \frac{V_{i,j}-V_{i+1,j}}{dx} h_{i+\frac{1}{2},j} dy +
                 \frac{V_{i,j}-V_{i-1,j}}{dx} h_{i-\frac{1}{2},j} dy, 
  \label{EqFinDiff}
\end{split}
\end{equation}

except for boundaries and corners of the domain,  where one or two
terms drop from the right hand side.  In this equation, $\sigma$ is
fixed by the material, $dx$ and $dy$ by the  geometry, currents and
voltages are measured, and the heights are the unknowns. The substitution  
$2/h_{i,j+\frac{1}{2}}=1/h_{i,j}+1/h_{i,j+1}$ and similar ones 
were omitted for brevity. Here the adoption of an harmonic mean instead of 
an arithmetic average is justified by the fact that the resistance 
between electrodes $i,j$ and $i,j+1$ is the sum of the resistances of the 
layers in between, which are inversely proportional to the local heights. 
Also note that, by evaluating the heights in the mid-points between electrodes, 
rather than at the electrodes, there would be a total of $2mn-m-n$ unknown 
heights. 

Eq.\ref{EqFinDiff} describes a set of $mn$ equations in $mn$ unknowns $h_{ij}$.  
The problem can be cast in matricial form $\bf I =A
h$. Here $\bf I$ and $\bf h$  are one-dimensional arrays containing
$mn$ values of currents and  heights, respectively. The
block-diagonal matrix $\bf A$ is large, but features  only five
non-vanishing elements in each row (or four, or three, in rows
corresponding to boundaries or corners), easily deduced  from
Eq.\ref{EqFinDiff}.  Ultimately we can solve for the LM heights by a
simple matrix inversion,  ${\bf h}= {\bf A}^{-1} {\bf I}$.

It should be noted that Ohm's law 
${\bf j} = \sigma {\bf E}$ was used in Eq.\ref{EqFinDiff}, 
instead of the more general 
${\bf j} = \sigma( {\bf E}  + {\bf v} \times {\bf B})$. This is legitimate under 
the assumptions that: (1) the liquid wall is thin and not significantly 
bulging or depleting ($v_z=0$) and (2) there is no error field orthogonal to the 
wall ($B_z=$0). Under these assumptions, ${\bf v} \times {\bf B}$ has no 
$x$ nor $y$ components, hence it cannot perturb the 
currents between electrodes. An alternative requirement is that (3) the flow 
is slow enough that the $x$ and $y$ components of 
${\bf v} \times {\bf B}$ are negligible compared with the corresponding components 
of $\bf E$. A realistic system violates these assumptions, and 
Eq.\ref{EqFinDiff} needs to be generalized:
\begin{equation}
\begin{split}
  \frac{I_{ij}}{\sigma} 
  =& \left[ \frac{V_{i,j}-V_{i,j+1}}{dy} +(v_xB_z-v_zB_x)_{i,j+\frac{1}{2}}
     \right] h_{i,j+\frac{1}{2}} dx \\
  +& \left[ \frac{V_{i,j}-V_{i,j-1}}{dy} -(v_xB_z-v_zB_x)_{i,j-\frac{1}{2}}
     \right] h_{i,j-\frac{1}{2}} dx \\
  +& \left[ \frac{V_{i,j}-V_{i+1,j}}{dx} +(v_zB_y-v_yB_z)_{i+\frac{1}{2},j}
     \right] h_{i+\frac{1}{2},j} dy \\
  +& \left[ \frac{V_{i,j}-V_{i-1,j}}{dx} -(v_zB_y-v_yB_z)_{i-\frac{1}{2},j}  
     \right] h_{i-\frac{1}{2},j} dy.  
  \label{EqFinDiffB}
\end{split}
\end{equation}

This requires knowledge of the components of $\bf v$ and $\bf B$ in the 
midpoints between electrodes. Such knowledge could be provided by separate 
diagnostics, assumptions or calculations, or $\bf v$ and $\bf B$ might be 
reasonably fixed in the experiment. With this external input to matrix $\bf A$, 
the problem can be solved as a simple matrix inversion again, 
${\bf h} = {\bf A}^{-1} {\bf I}$. 

\begin{figure}[t]
  \includegraphics[width=0.5\textwidth]{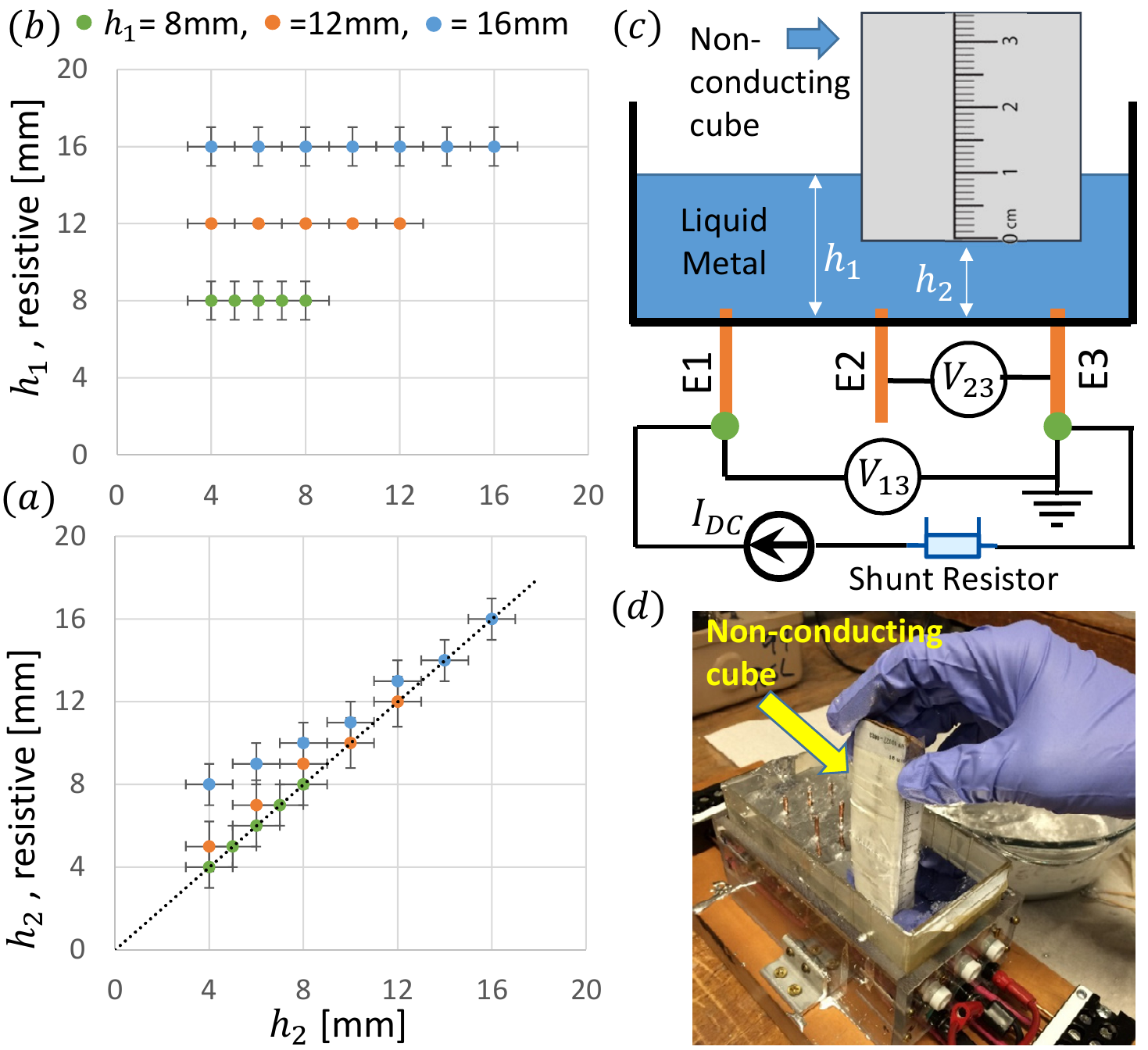}
  \caption{Scheme of the space-resolved measurement test and the results; a) and b) $h_2$ and $h_1$ resistively measured versus $h_2$ measured by a Teflon coated ruler, c) scheme of the sensor, with the applied current and resistance measurement and d) experimental setup.}
  \label{Three_Electrodes_Test}
\end{figure}

\subsection{Experiment}
Three 1 mm tall copper electrodes (one column of the $3\times 4$
matrix) were used for space-resolved measurements. The DC
current generator is connected to the first and the last electrodes in
the row (Fig. \ref{Three_Electrodes_Test}). 
The LM thickness between electrodes
E2 and E3 is varied by inserting a non-conducting cube in the space
over and between these electrodes. Heights $h_1$ and $h_2$ were
measured simultaneously by the resistive sensors, using 
Eq.\ref{EqFinDiff} for $m$=3 and $n$=1. 
The results are plotted in Fig. \ref{Three_Electrodes_Test}. 
The sensor is more accurate when measuring lower
$h_1$ values. This is reasonable, since by increasing the LM height,
the current between the short electrodes is not uniformly distributed,
which decreases the accuracy of the theoretical model. 

Another test was accomplished with the same setup, where this time the
LM thickness was varied by tilting the entire LM pot.
A difficulty was observed when resistively measuring the increase of
$h_1$. The sensor was not able to follow the increase, while it was
correctly measuring the decrease in $h_2$. The specular issue
occurred when tilting to the opposite direction.

\section{Time-resolved measurements}\label{SecTime}
Voltages at and currents between $3\times1$ electrodes 
were digitized at $10^5$ kSa/s 
and analyzed with Eq.\ref{EqFinDiff}.   
This provided space- and time-resolved information on LM thickness, as this 
was being periodically perturbed by a platform shaker oscillating by 
$\pm$10 mm at 0-3.3 Hz. The introduction of these flows was 
important because LM walls will flow in a reactor, and 
be subject to secondary flows due to instabilities, turbulence and 
other effects \cite{Narula,Morley,Jaworski,Ait,TahaPPCF}. The method presented 
is general with respect to the type of flow.  
 
\begin{figure}[t]
  \includegraphics[width=0.5\textwidth]{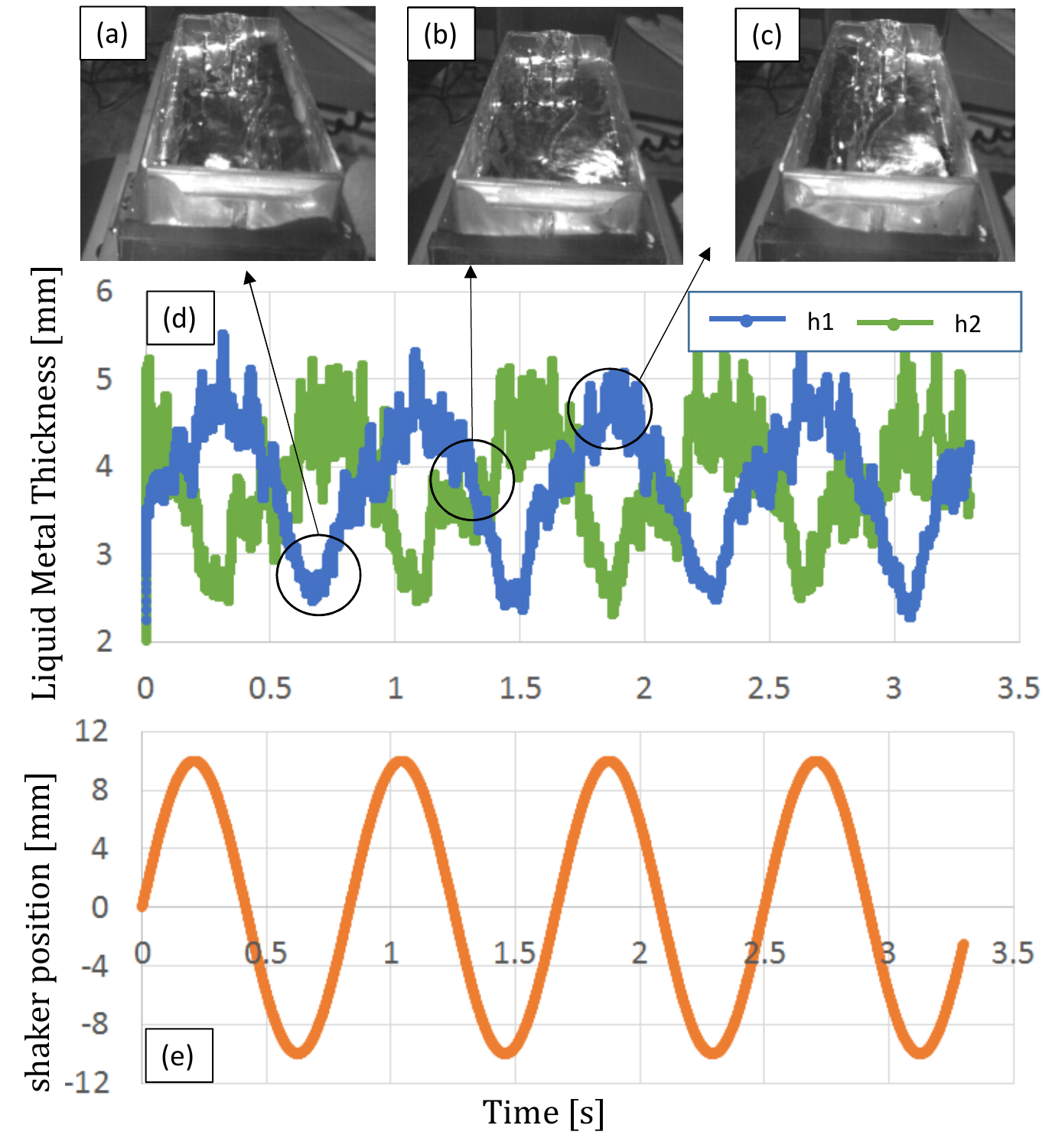}
  \caption{Results of the space- and time-resolved measurements using a matrix of 3$\times$1 copper, 1mm tall electrodes. The sensor container, including 4 mm deep of Galinstan is linearly moving in the horizontal direction at 72 RPM. a), b) and c) show the LM sensor at three different time steps corresponding to the points indicated on the measurement graph (d) and e) is the shaker position. $h_1$ and $h_2$ correspond to the heights measured by the left and right sensors.}
  \label{TimeResol}
\end{figure}

The only issue was that currents had to cross thick, short pieces of 
a good conductor (Galinstan). The resistances of 
interest were therefore small, the voltage signals were also small, 
and the signal-to-noise ratio relatively low. Still, 
despite noise, it was possible to achieve the desired precision of 1 mm and 
exceed the desired time-resolution (\dots instead of 10 ms), as illustrated by 
Fig.\ref{TimeResol}. Higher precisions are obviously achievable for coarser 
time resolutions.

Note that, due to the finite width of the container, shallow liquid metal and 
large oscillatory motion, the surface waves excited in the LM were highly 
non-linear. Therefore, it was not surprising that, while periodic, the 
time-traces in Fig.\ref{TimeResol} were not pure sine-waves. 
Fast camera images confirmed the non-linear behavior and agreed on the 
existence of ``steps'' (Fig.\ref{TimeResol}).

\section*{Summary and conclusions}
Resistive measurements of liquid metal 
thickness were spatially and temporally resolved for the first time. 
A theory was developed for 
$m\times n$ electrodes and experimentally demonstrated for 
3$\times$1 electrodes.  
Measurements were carried out with Galinstan in the absence of plasma, 
but are expected to succeed 
also with Lithium, whose conductivity is nearly identical. 

Future work will be carried out in the presence of plasma. 
The diagnostic of thickness might require information from flowmeters and 
magnetics, due to complications associated with error fields and rapid flows, 
theoretically discussed in Sec.\ref{SecSpace}.

\end{document}